\documentclass[aps,preprint,superscriptaddress]{revtex4}

\usepackage{epsfig,color}
\usepackage{amsmath}

\newcommand{\pderiv}[2]{\frac{\partial #1}{\partial #2}}

\newcommand{\infint}{\int \limits_{-\infty}^{\infty}}
\newcommand{\half}{\frac{1}{2}}
\newcommand{\e}[1]{\text{e}^{#1}}
\definecolor{red}{rgb}{1,0,0}

\newcommand{\eq}[1]{Eq.~(\ref{#1})}

\begin{document}

\title{$q$-Gaussians in the porous-medium equation: stability and 
time evolution}

\vskip \baselineskip

\author{Veit Schw\"ammle}
\thanks{E-mail address: veit@cbpf.br}
\affiliation{Centro Brasileiro de Pesquisas F\'isicas, 
Rua Xavier Sigaud 150, Rio de Janeiro, RJ 22290-180, Brazil}

\author{Fernando D. Nobre}
\thanks{Corresponding author: E-mail address: fdnobre@cbpf.br}
\affiliation{Centro Brasileiro de Pesquisas F\'isicas, 
Rua Xavier Sigaud 150, Rio de Janeiro, RJ 22290-180, Brazil}

\author{Constantino Tsallis}
\thanks{E-mail address: tsallis@cbpf.br}
\affiliation{Centro Brasileiro de Pesquisas F\'isicas, 
Rua Xavier Sigaud 150, Rio de Janeiro, RJ 22290-180, Brazil}
\affiliation{Santa Fe Institute,
1399 Hyde Park Road, 
Santa Fe, New Mexico 87501, USA}

\date{\today}

\begin{abstract}
The stability of $q$-Gaussian distributions as particular solutions of the
linear diffusion equation and its generalized nonlinear form, 
$\pderiv{P(x,t)}{t} = D \pderiv{^2 [P(x,t)]^{2-q}}{x^2}$, 
the \emph{porous-medium equation}, is investigated through both numerical 
and analytical approaches. It is shown that an \emph{initial} $q$-Gaussian,
characterized by an index $q_i$, approaches the
\emph{final}, asymptotic solution, characterized by an
index $q$, in such a way that the relaxation rule for the kurtosis
evolves in time 
according to a $q$-exponential, with a \emph{relaxation} index 
$q_{\rm rel} \equiv q_{\rm rel}(q)$.
In some cases, particularly when one attempts to transform an
infinite-variance distribution ($q_i \ge 5/3$) into a finite-variance 
one ($q<5/3$), 
the relaxation towards the asymptotic solution may occur very
slowly in time. This fact might shed some light on the slow relaxation, for
some long-range-interacting many-body Hamiltonian systems, from
long-standing quasi-stationary states to the ultimate thermal equilibrium
state. 

\vskip \baselineskip

\end{abstract}

\maketitle

\section{Introduction}

The linear diffusion equation, which is one of the most important  
differential equations of classical physics, rules the time evolution of the
probability distribution associated with a particle diffusing in a
homogeneous medium. This probability distribution spreads in time, with
its second moment increasing linearly with time, 
being appropriate for a description of many physical phenomena, usually
classified as normal diffusion. Such an equation presents a structure very
common in physics, e.g., it is analogous to the heat-conduction equation,
and by introducing an additional term, characterizing an external harmonic
force field, it becomes the linear Fokker-Planck 
equation (FPE)~\cite{VanKampen:81,Risken:89}. The linear FPE is
essentially associated with the Boltzmann-Gibbs (BG) formalism, in the
sense that the Boltzmann distribution, which is
usually obtained through the maximization of the BG entropy under 
certain constraints, also appears as the 
stationary solution of the linear FPE~\cite{Risken:89,Cover:91}.
However, such simple linear equations are not appropriate for dealing with
many physical 
situations, characterized by anomalous diffusion, like particle transport
in disordered media \cite{Muskat:37} and motion in optical lattices
\cite{Lutz:03}.  

Due to the recent advance in computer technology, some problems in physics
that remained unexplored for a long time are now under investigation, at
least from the computational point of view. In particular, one should single
out those problems described in terms of nonlinear differential equations,
which have led 
to new features and interesting puzzles that keep challenging many
physicists nowadays. The nonlinear FPEs~\cite{Frank:05}, which
are formulated by introducing modifications in the standard FPE,  
appear naturally as good candidates for describing anomalous-transport
processes. In most cases, the nonlinear FPEs are proposed as simple
phenomenological generalizations of the linear FPE
\cite{Plastino:95,Tsallis:96,Borland:98b,Frank:99}, although it is possible to
obtain them through approximations in the master equation
\cite{Curado:03,Nobre:04,Schwaemmle:07c}. Recently, a relation involving
quantities of the FPE and entropic forms was proposed, in a proof of the  
H-theorem using nonlinear FPEs \cite{Schwaemmle:07c,Schwaemmle:07b};  
as a consequence of such a relation, one obtains that the BG entropy is
directly connected to the linear FPE, whereas generalizations of the BG
entropy are associated with nonlinear FPEs 
\cite{Schwaemmle:07c,Schwaemmle:07b}. This
reinforces the belief that nonlinear FPEs are intimately related to
nonextensive statistical mechanics 
\cite{next:04,next:05}. 

If one considers the nonlinear 
FPE associated with the nonadditive entropy $S_{q}$ \cite{Tsallis:88}
in the
absence of an external force field, one gets \cite{Plastino:95,Tsallis:96} 
the porous-medium equation (also known as nonlinear heat 
equation)~\cite{Vazquez:06},

\begin{equation}
  \label{eq:NDiffEq}
  \pderiv{P(x,t)}{t} = D \pderiv{^2 [P(x,t)]^{2-q}}{x^2}~,
\end{equation}

\vskip \baselineskip
\noindent
which governs the time evolution of the probability distribution
$P(x,t)$ for finding a diffusing particle in the position $x$ at time
$t$, in a medium characterized by a diffusion constant $D$; 
the linear diffusion equation comes out as a particular case, by
considering $q=1$. 
It should be noticed that, within a numerical analysis, as will be one of
the main purposes of the present work, the parameter $q$ above
corresponds to the index associated with the asymptotic behavior of the 
solution of \eq{eq:NDiffEq}, known as a $q$-Gaussian, related to
the nonadditive entropy \cite{Tsallis:88}.

Recently, classical inertial long-range-interaction Hamiltonian systems
have attracted a lot of attention
\cite{LatoraRuffo:98,Anteneodo:98,Moyano:06,Pluchino:07,Pluchino:08,%
nobretsallis:03,nobretsallis:04,Ruffo:07,Ruffo:07b}. These systems 
consist in
assemblies of classical rotators, which evolve in time, e.g., in a
plane 
(XY rotators \cite{LatoraRuffo:98,Anteneodo:98,Moyano:06,Pluchino:07,%
Pluchino:08}), 
or in a sphere (Heisenberg rotators \cite{nobretsallis:03,nobretsallis:04}).  
For the case of infinite-range ferromagnetic interactions, i.e., in the 
mean-field limit, one may calculate thermodynamic properties analytically,
within the BG canonical ensemble, and in particular, verify the 
existence of a continuous phase transition. 
The interesting aspect about these systems is that one may compute their
time evolution through a direct integration of their equations of motion,
without introducing \emph{a priori} any phenomenological
dynamic rules, but just using Newton's law. A quite curious behavior has
shown up by starting the 
numerical-integration procedure  
with initial conditions very different from those required in the standard
BG equilibrium: the system gets trapped in metastable states, before approaching
their corresponding terminal thermal equilibria. These metastable states
are  
characterized by ``kinetic temperatures'' that are different from the
equilibrium ones; besides that, the duration of such states increases with
the number of rotators, $N$. 
Hence, if one considers the 
thermodynamic limit ($N \rightarrow \infty$) before the long-time 
limit, these systems will remain in these metastable states and will
never reach their terminal equilibrium state, in such a way that  
the phase space will not be equally and completely covered, i.e., these
systems are nonergodic.  
Moreover, in such metastable states,  
the maximum Lyapunov exponent approaches zero, as 
$N \rightarrow \infty$ \cite{Anteneodo:98,nobretsallis:03}, 
contrary to what is expected in a standard BG 
equilibrium state. 
For finite values of $N$, at large enough -- but
realizable computational times -- one approaches a state that is presumably
the terminal thermal equilibrium state, in the sense that its kinetic
temperature is in agreement with the one obtained through the
BG canonical-ensemble calculations.  
However, although the kinetic temperature of  
the long-time limit state agrees with the one obtained from the
canonical-ensemble calculations, other properties may still not coincide
with those expected in a true equilibrium BG state; as an example, one has
a recent analysis of the angles described by the infinite-range-interaction
XY rotator model, for which their distribution in the long-time-limit is
well-fitted by a $q$-Gaussian, with $q\approx1.5$~\cite{Moyano:06}. This
suggests that the BG equilibrium state is approached through different
steps, one of them being the attainment of the equilibrium temperature; a
relevant question concerns how long will it take for the system to reach
completely the final BG equilibrium state. 

In the present work we search for clues on how the approach to equilibrium
occurs in the above-mentioned Hamiltonian models, by investigating the time
evolution of probability distributions in a much simpler system, i.e., 
the porous-medium equation. For that, we integrate 
\eq{eq:NDiffEq}, starting the integration procedure with 
an initial distribution different from its asymptotic solution.
In particular, we will consider as initial distribution a
$q$-Gaussian characterized by an
entropic index $q_{i}$ ($q_{i} \neq q$), and will follow the time
evolution of such a distribution towards the final
(i.e., asymptotic) $q$-Gaussian,  
specified by the index $q_{f}$, that hopefully, $q_{f} \equiv q$. 
Exploring the stability of $q$-Gaussians in an environment
given by \eq{eq:NDiffEq} may help to understand why metastable states
that appear in the 
infinite-range-interaction models of rotators remain stable
over such long periods. 
In the following section we discuss the exact solutions of the 
porous-medium equation and their connection to anomalous diffusion. In
Sec.~\ref{sec:moments} we introduce generalized moments, as well as 
a generalized kurtosis, which are more appropriate for dealing with
fat-tailed distributions. 
In Sec.~\ref{sec:q1} we analyze the time evolution of a probability
distribution by following the linear diffusion
equation [$q=1$ in \eq{eq:NDiffEq}], provided that the initial distribution
is given by a  
$q$-Gaussian with $q_i \neq 1$.
In Sec.~\ref{sec:qfneq1} we carry a similar analysis for the 
porous-medium equation [$q \neq 1$ in \eq{eq:NDiffEq}], 
having as initial state a $q$-Gaussian with $q_i \neq q$. 
In Secs.~\ref{sec:q1} and~\ref{sec:qfneq1} we show, by 
monitoring the time evolution of the kurtosis, that the approach to the
final $q$-Gaussian obeys a $q$-exponential function, 
characterized by a relaxation index $q_{\rm rel}$. At this point, it is
important to stress that in the present work we deal, in principle, with
four indexes: (i) the index $q$ defined by \eq{eq:NDiffEq}; (ii) $q_{i}$, 
associated with the initial $q$-Gaussian distribution; (iii) $q_{f}$,
associated with the final $q$-Gaussian distribution (for which one expects,
$q_{f} \equiv q$); (iv) $q_{\rm rel}$, related to the $q$-exponential of the 
relaxation towards the asymptotic distribution. 
However, in the present work we have found no evidence, either in our
analytical or numerical approaches, of $q_f \neq q$; hence, we shall
assume from now on that $q_f \equiv q$. Therefore, we will restrict our
analysis 
to three indexes, namely, $q, \, q_i$, and $q_{\rm rel}$, as defined above.
Finally, in the last section we present our main conclusions. 
 
\section{Exact solutions of the porous-medium equation}
\label{sec:part}

In this section we discuss briefly the well-known exact solutions for a
diffusing particle following \eq{eq:NDiffEq}. In order to guarantee 
the preservation of the normalization for all times $t$, one should impose
the probability distribution, 
together with its first derivative to be zero at infinity, 

\begin{equation}
P(x,t)|_{x \rightarrow \pm \infty} = 0~; \quad
\left. {\partial P(x,t) \over \partial x} 
\right|_{x \rightarrow \pm \infty} = 0~,  
\quad (\forall t)~. 
\label{eq:p_inf}
\end{equation}

\vskip \baselineskip
\noindent
If one chooses a perfectly localized particle as the 
initial state, $P(x,0) = \delta(x_0)$ 
[$\delta(x)$ denotes the delta-function], then, 
following Refs.~\cite{Plastino:95,Tsallis:96}, one can write the 
solution of  
\eq{eq:NDiffEq}, satisfying the conditions of \eq{eq:p_inf}, in terms 
of a $q$-Gaussian, 

\begin{equation}
\label{eq:Diff_Ansatz}
P(x,t) = 
Z_{q} b_{q}(t) \ \e{- b_{q}^2(t) (x-x_0)^2}_{q}~, \quad 
\text{for  } q<3~, \\
\end{equation}

\vskip \baselineskip
\noindent
where $\e{x}_q=\left[ 1+(1-q) x\right]_+^{1/(1-q)}$ 
(herein, the 
bracket $[C]_+=C$, for $C\geq 0$, and is zero otherwise)
represents 
the $q$-generalization of the standard
exponential function that is recovered in the 
limit $q \rightarrow 1$; its inverse, known as the 
$q$-logarithmic function, is given by
$\ln_{q} x=(x^{1-q}-1)/(1-q)$.
This solution presents a compact support for $q<1$, and 
exhibits power-law tails for $q>1$. 
The time-dependent part of the solution, $b_{q}(t)$, and the
normalization constant, $Z_{q}$, are given, respectively, by

\begin{align}
\label{eq:NDiffb}
b_{q}(t) = \left[ 2 D (2-q) (3-q) Z_q^{1-q} \ t
\right]^{1/(q-3)}~, \quad (D(2-q)>0)~,
\end{align}

\vskip \baselineskip
\noindent

\begin{equation}
\label{eq:Diff_Z}
Z_{q} = 
\begin{cases}
\sqrt{\frac{|D|(q-1)}{\pi}}~ \frac{\Gamma \left( \frac{1}{q-1} \right)}
{\Gamma \left(
\frac{1}{q-1} -\half \right)}~,  \quad & \text{for } 1<q<3~, \\
\sqrt{\frac{D}{\pi}}~,  \quad & \text{for }  q=1~, \\
\sqrt{\frac{D(1-q)}{\pi}}~\frac{ \Gamma \left( 1 + \frac{1}{1-q} \right)}
{\Gamma \left( \frac{3}{2}
+ \frac{1}{1-q} \right)}~, \quad  & \text{for }  q<1~, 
\end{cases}
\end{equation}

\vskip \baselineskip
\noindent
where $\Gamma(x)$ represents the Gamma-function.

The diffusion is usually characterized by the time behavior of the second
moment of the distribution, which is given by 
$<x^2> = (b_{q}(t))^{-2}$, scaling as $t^{2/(3-q)}$. 
Hence, $q=1$ yields a linear increase in time, i.e., normal diffusion, 
whereas \eq{eq:NDiffEq} leads to anomalous diffusion for 
$q\neq1$. Within
anomalous diffusion, one may distinguish 
super-diffusion ($q>1$), characterized by long-tailed distributions, from
sub-diffusion ($q<1$), related to compact-support distributions. 

If one starts the numerical integration of \eq{eq:NDiffEq}
with $P(x,0) = \delta(x_0)$, one follows the corresponding   
$q$-Gaussian, associated with $<x^2> \sim t^{2/(3-q)}$.
However, if one uses as an initial distribution a 
$q$-Gaussian specified by an index $q_i \neq q$, 
the numerical procedure will
take some time to gradually change from such an initial, to the 
final $q$-Gaussian
distribution. Herein, we will be
particularly interested in measuring 
the time that the system takes to approach its final distribution
asymptotically. Therefore, different initial states of the system are
expected to yield different relaxation behavior. In the analysis that
follows, the solution presented in 
Eqs.~(\ref{eq:Diff_Ansatz})--(\ref{eq:Diff_Z}) will be our reference and
its kurtosis, to be introduced in the next section, will be an important
quantity to characterize the relaxation behavior. 

\section{Generalized moments and kurtosis}
\label{sec:moments}

The kurtosis is usually defined in terms of the ratio between the
fourth, and the square of the second moments of a given distribution.
However, for the $q$-Gaussian distribution defined in the previous section,
one gets divergences in its even moments, in such a way that
the second moment diverges for $q \geq 5/3$, whereas the fourth moment diverges
for $q \geq 7/5$. Therefore, the standard definitions of moments and kurtosis
become useless for certain ranges of $q$ values. In the present 
section, we introduce generalized moments, and
apply them in a definition of a generalized kurtosis.  

Let us define the generalized $n$-th moment of a given distribution as, 

\begin{equation}
\label{eq:Mom_n}
<x^n>_r~=~{\infint dx ~x^n [P(x,t)]^{r} \over \infint dx [P(x,t)]^{r}}~,
\end{equation}

\vskip \baselineskip
\noindent
where $n$ is a positive integer and $r \geq 0$. For the $q$-Gaussian
distribution, defined in the previous section, one has 

\begin{align}
\label{eq:Moms}
<x^2>_r = & 
\begin{cases} 
\half \frac{b_{q}^{-2}(t)}{q-1} \frac{\Gamma \left(\frac{r}{q-1}
-\frac{3}{2} \right)}{\Gamma \left( \frac{r}{q-1}-\half \right)}~,  
& \text{for } 0<(q-1)<\frac{2}{3}r~, \\
\half \frac{b_1^{-2}}{r}~, & \text{for } q=1~, \\
\half \frac{b_{q}^{-2}(t)}{1-q} \frac{\Gamma \left(\frac{r}{1-q}
+\frac{3}{2} \right)}{\Gamma \left( \frac{r}{1-q}+\frac{5}{2} \right)}~,  
& \text{for } q<1~, 
\end{cases}
\\
\label{eq:Moms2}
<x^4>_r =   & 
\begin{cases}
\frac{3}{4} \frac{b_{q}^{-4}(t)}{(q-1)^2} \frac{\Gamma 
\left(\frac{r}{q-1}-\frac{5}{2} \right)}{\Gamma \left( \frac{r}{q-1}
-\half \right)}~, & \text{for } 0<(q-1)<\frac{2}{5}r~, \\
\frac{3}{4} \frac{b_1^{-4}}{r^2}~, & \text{for } q=1~, \\
\frac{3}{4} \frac{b_{q}^{-4}(t)}{(1-q)^2} \frac{\Gamma 
\left(\frac{r}{1-q}+\frac{3}{2} \right)}{\Gamma \left( \frac{r}{1-q}
+\frac{7}{2} \right)}~, & \text{for } q<1~.   
\end{cases}
\end{align}

\vskip \baselineskip
\noindent
Considering the above moments, one may define the following generalized
kurtosis, 

\begin{equation}
\label{eq:kurtosis}
\kappa_{r,s}(q) = \frac{<x^4>_r}{(<x^2>_s)^2} = 
\begin{cases}
3 \frac{\Gamma\left(\frac{r}{q-1}-\frac{5}{2}\right) 
\left(\Gamma\left(\frac{s}{q-1}-\frac{1}{2}\right) \right)^2}
{\Gamma\left(\frac{r}{q-1}-\frac{1}{2} \right) 
\left(\Gamma\left(\frac{s}{q-1}-\frac{3}{2}\right) \right)^2}~,  
& \text{for } 0<(q-1)< \text{min} \left( \frac{2}{3}r ,  \ \frac{2}{5}s
\right)~, \\
3 \frac{s^2}{r^2}~, & \text{for } q=1~, \\
3 \frac{\Gamma\left(\frac{r}{1-q}+\frac{3}{2} \right) 
\left(\Gamma\left(\frac{s}{1-q}+\frac{5}{2}\right) \right)^2}
{\Gamma\left(\frac{r}{1-q}+\frac{7}{2} \right) 
\left(\Gamma\left(\frac{s}{1-q}+\frac{3}{2}\right) \right)^2}~,
& \text{for } q<1~, 
\end{cases}
\end{equation}

\vskip \baselineskip
\noindent
If one uses the property $\Gamma(x+1)=x\Gamma(x)$, it is possible to
write this kurtosis in a form that covers all three possibilities
above,

\begin{equation}
\label{eq:Kurt_fin1}
\kappa_{r,s}(q)= 3 \ \frac{(2s-3(q-1))^2}{(2r-3(q-1))(2r-5(q-1))}~, 
\quad  \left[(q-1)< \text{min} 
\left( \frac{2}{3}r ,  \ \frac{2}{5}s \right) \right]~.  
\end{equation}

\vskip \baselineskip
\noindent
In the numerical integration of \eq{eq:NDiffEq} the  
\emph{initial} $q$-Gaussian, characterized by an
entropic index $q_{i}$ ($q_{i} \neq q$), will evolve in time towards
the \emph{final} $q$-Gaussian. 
Obviously, it is 
desirable to have $\kappa_{r,s}(q)$ always finite when the parameter $q$ 
varies in the interval $q_{i} \rightarrow q$; unfortunately, this is
not possible in some cases, and for this purpose, one has
to choose the exponents $r$ and $s$ conveniently. Herein, we choose these
exponents by giving preference to a finite kurtosis in the  
asymptotic limit ($t\gg1$).

\begin{figure}[tb]
\centering
\includegraphics[angle=270,width=0.60\textwidth]{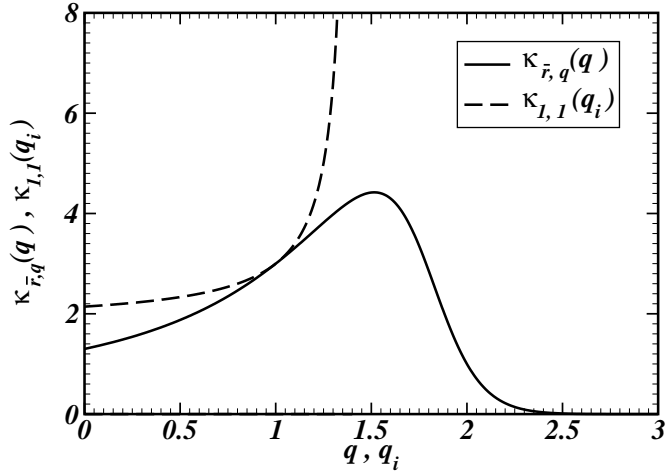}
\caption{
The kurtosis at the beginning of the numerical procedure
[\eq{eq:Kurt_gen_s}], characterized by
linear diffusion in the asymptotic regime ($q=1$), $\kappa_{1,1}(q_i)$,
is exhibited versus $q_i$ (dashed line). Also shown is the kurtosis 
at the final regime, $\kappa_{\bar{r},q}(q)$ 
versus $q$ (full line). Notice that the standard Gaussian value,
$\kappa_{1,1}(1)=3$, is recovered from both cases.}
\label{fig:kurtosis_fin}
\end{figure}

In a recent proof of a generalized central-limit theorem, $q$-Gaussian
distributions appear as a result of given composition 
rules~\cite{Umarov:06,Vignat:07}. Inspired by some results of this theorem,
we will consider the choices
$s=q$ and $r = \bar{r} \equiv (q+1)/(3-q)$; the choice 
$r = \bar{r}$ ensures a finite second moment in the asymptotic limit 
($t\gg1$). Substituting these quantities in \eq{eq:Kurt_fin1}, one gets the
following expression for the kurtosis at the beginning of the numerical
procedure, 

\begin{equation}
\label{eq:Kurt_gen_s}
\kappa_{\bar{r},q}(q_i)=3 \ \frac{[3-q]^2[2q-3(q_i-1)]^2}
{[11-q-3q_i(3-q)][17-3q-5q_i(3-q)]}~,
\end{equation}

\vskip \baselineskip
\noindent
which is finite, provided that
$(q_i-1)< \text{min} \left[(2/5)(q+1)/(3-q),(2/3)q \right]$. 
An interesting particular case of \eq{eq:Kurt_gen_s} is the one
characterized by linear diffusion in the asymptotic regime, i.e., 
$q=1$, yielding 
$\kappa_{1,1}(q_i)=3 \cdot (5-3q_i)/(7-5q_i)$, which leads to a divergence
at $q_i=7/5$, as shown in Fig.~\ref{fig:kurtosis_fin}. 

The same choices for the exponents $r$ and $s$ yield the kurtosis of
\eq{eq:Kurt_fin1} in the asymptotic regime, 

\begin{equation}
\label{eq:kurt_q}
\kappa_{\bar{r},q}(q) = 3 \ \frac{(3-q)^4}
{(3q^2-10q+11)(5q^2-18q+17)}~. 
\end{equation}

\vskip \baselineskip
\noindent
In Fig.~\ref{fig:kurtosis_fin} we exhibit the kurtosis above versus $q$,
showing that, as expected, it does not diverge. 

Therefore, when the $q$-Gaussian changes between the two entropic indices 
$q_{i} \rightarrow q$, the kurtosis evolves in time, changing its 
behavior between those described in Eqs.~(\ref{eq:Kurt_gen_s}) and 
(\ref{eq:kurt_q}), respectively; although the kurtosis may be infinite 
during its time evolution, it will be finite in the asymptotic limit.

\section{$q=1$ and arbitrary initial distributions}
\label{sec:q1}

In this section we restrict our study to the linear diffusion equation
[\eq{eq:NDiffEq} with $q=1$], 
analyzing the time evolution of 
different initial states 
given by $q$-Gaussians [cf. \eq{eq:Diff_Ansatz}], characterized by
distinct entropic indexes $q_i$. 

\begin{figure}[tb]
\centering
\includegraphics[angle=270,width=0.60\textwidth]{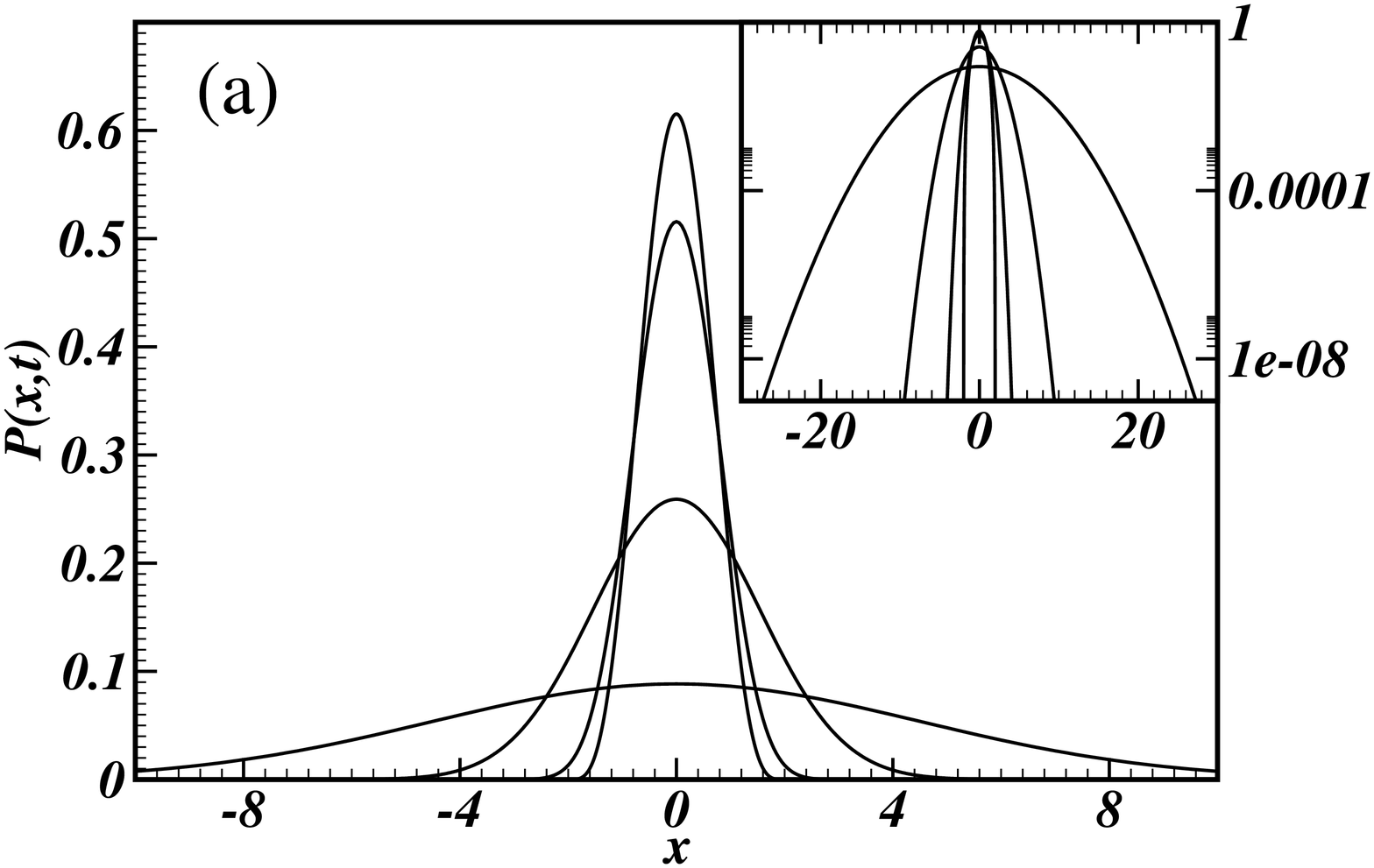}
\includegraphics[angle=270,width=0.60\textwidth]{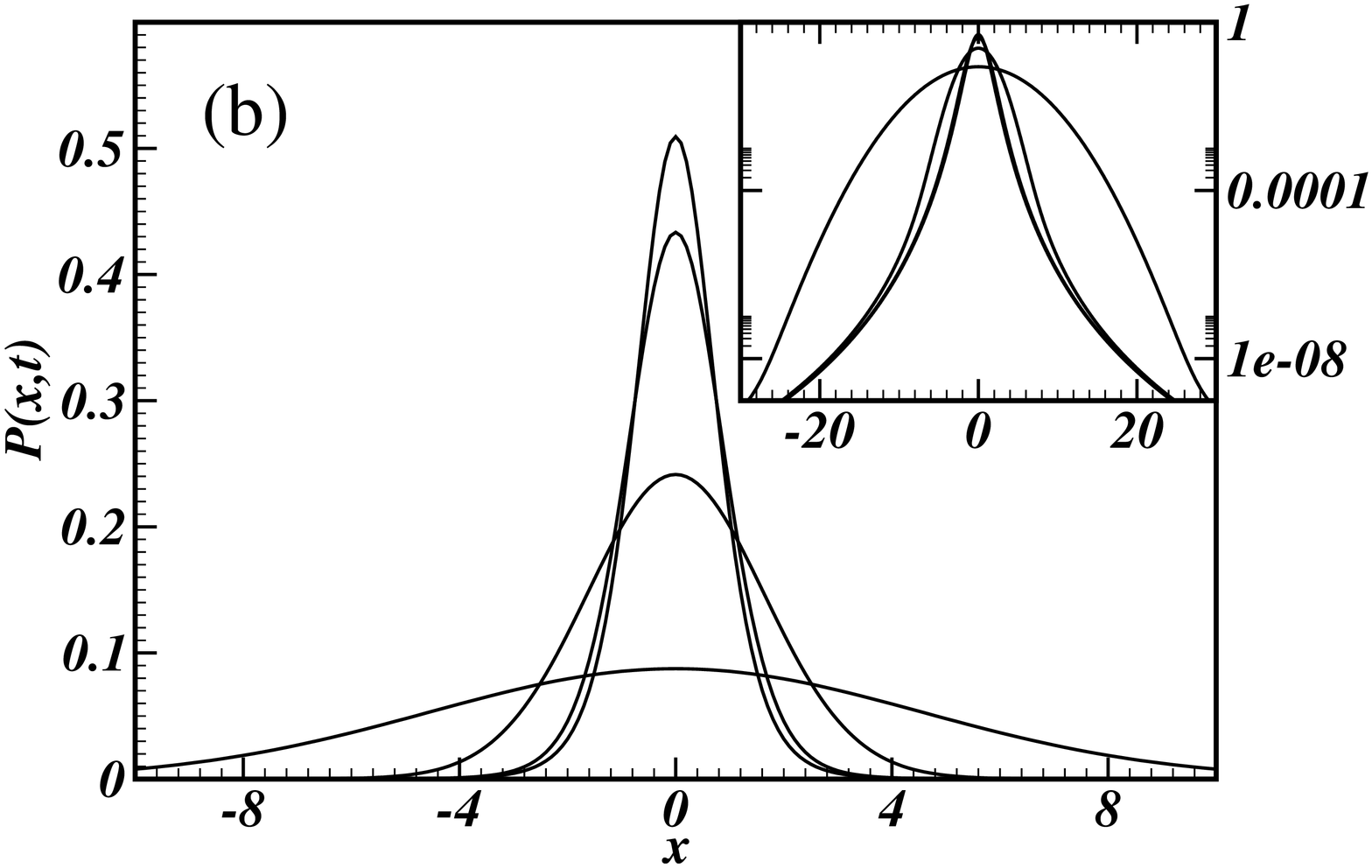}
\caption{Time evolution of the probability distribution associated with the
linear diffusion equation at typical times,
$t=0,0.1,1,10$ (from top to bottom). The system is initialized with 
$q$-Gaussians, characterized by initial entropic indexes $q_i=0.75$ (a)
and $q_i=1.25$ (b). In the insets we exhibit the same distributions on a 
semi-logarithmic scale.}
\label{fig:DistrQ1}
\end{figure}

As an illustration, we exhibit in Fig~\ref{fig:DistrQ1} the time evolution
of $P(x,t)$, 
starting the numerical integration with two typical $q$-Gaussians, namely, 
$q_i=3/4$ [Fig~\ref{fig:DistrQ1}(a)] and $q_i=5/4$
[Fig~\ref{fig:DistrQ1}(b)]. 
On the linear scale, one observes no notable difference between the
time evolution of these probability distributions. Nevertheless, on the
log-linear scale (cf. insets of Fig.~\ref{fig:DistrQ1}), one sees clearly
that the initial distribution relaxes faster to the Gaussian limit in the
case $q_i=3/4$, whereas the fat tails
remain stable over a longer period for $q_i=5/4$. In particular, in this
later case, one notices the presence of an inflection point, characteristic
of $q$-Gaussians with $q>1$, at intermediate times (i.e., in the transient
regime), which disappears when the distribution approaches the Gaussian
limit.
  
In what concerns the kurtosis, $\kappa_{1,1}(q_i)$, introduced in the
previous section, it may be calculated analytically, for $q_i<7/5$. This
quantity enables a
measure of the time the system needs to reach asymptotically the Gaussian
distribution; from now on, we will indicate its time dependence explicitly,
by referring to it as 
$\kappa_{1,1}(q_i,t)$. As we deal here with a linear partial differential
equation, 
the calculation of the moments may be carried out exactly by the use of the
Green's function method. 
The final distribution is a standard Gaussian, 

\begin{equation}
\label{eq:Gauss_sol}
P(x,t) = \frac{1}{2 \sqrt{D \pi t}} 
\exp\left[ -\frac{(x-x_0)^2}{4Dt} \right]~, 
\end{equation}

\vskip \baselineskip
\noindent
and the corresponding Green's function is given by~\cite{Mathews:70}, 

\begin{equation}
\label{eq:Green}
G(y,x|t) = \frac{1}{2 \sqrt{\pi t}} 
\exp\left[{-\frac{(x-y)^2}{4D t}} \right]~,
\end{equation}

\vskip \baselineskip
\noindent
from which one obtains the time-dependent solution of the diffusion
equation, 

\begin{equation}
\label{eq:Green_sol}
P(x,t) = \infint G(y,x|t) P(y,0) dy~,
\end{equation}

\vskip \baselineskip
\noindent
for arbitrary initial distributions $P(x,0)$.
Let us now calculate the $n$-th moment of the time-dependent solution,

\begin{align}
\nonumber
<x^n>_{1} & = & \infint dx~ x^n \infint dy~ G(y,x|t) P(y,0) = 
\infint dy P(y,0) \infint dx ~x^n G(y,x|t)\\
\label{eq:moments} 
& & = \frac{1}{2 \sqrt{\pi t}} \infint dy~P(y,0) \infint dx~(x-y)^n  
\exp\left({-\frac{x^2}{4D t}} \right)~.
\end{align}

\vskip \baselineskip
\noindent
Using \eq{eq:moments}, we obtain the time evolution of the
moments of  $P(x,t)$ for arbitrary initial functions $P(x,0)$, just by
calculating standard Gaussian integrals. In order to obtain the
kurtosis, we calculate the second and fourth moments, respectively,

\begin{equation}
\label{eq:second_mom}
<x^2>_{1} = \infint dy P(y,0) \left( y^2 + 2 t \right) = \bar{y^2} + 2t~,
\end{equation}

\vskip \baselineskip
\noindent
and 

\begin{equation}
\label{eq:Fourth_mom}
< x^4 >_{1} = \infint dy P(y,0) \left( y^4 + 12 t y^2 + 12 t^2 \right) = 
\bar{y^4} + 12 \bar{y^2} t + 12 t^2~, 
\end{equation}

\vskip \baselineskip
\noindent
where $\bar{y^2}$ and $\bar{y^4}$ denote the standard
second and fourth moments 
of the initial distribution $P(x,0)$. 
Using these results, the kurtosis becomes,

\begin{equation}
\label{eq:kurtosis_lin}
\kappa_{1,1}(q_i,t) = \frac{<x^4>_{1}}{(<x^2>_{1})^2} 
= \frac{\bar{y^4} + 12 t ( \bar{y^2} + t)}{(\bar{y^2}+2t)^2}~.
\end{equation}

\vskip \baselineskip
\noindent
Therefore, the kurtosis' asymptotic value, 
$\lim_{t \rightarrow \infty}\kappa_{1,1}(1,t)=3$, is
approached according to,

\begin{equation}
\label{eq:kurtosis_2}
\kappa_{1,1}(q_i,t)-3 = \frac{\bar{y^4} - 3 (\bar{y^2})^2}{(\bar{y^2}+2t)^2}~.
\end{equation}

\vskip \baselineskip
\noindent
The equation above yields the kurtosis
in terms of the second and fourth moments of initial $q$-Gaussians,
provided that $q_i<7/5$. 
Notice that $\kappa_{1,1}(q_i)-3$ is negative, for $q_i<1$, 
and positive, for $1<q_i<7/5$. 
Defining $b \equiv b_{q_i}(t=0)=$ constant, and
using Eqs.~(\ref{eq:Moms}) and~(\ref{eq:Moms2}), 

\begin{align}
\label{eq:Moms_Gauss}
& \bar{y^2} = \frac{b^{-2}}{5-3q_i}~,  \quad (q_i<5/3), \\
& \bar{y^4} = 3 \frac{b^{-4}}{(5-3q_i)(7-5q_i)}~, \quad (q_i<7/5), 
\end{align}

\vskip \baselineskip
\noindent
and thus,

\begin{align}
\label{eq:Kurtosis_3}
\kappa_{1,1}(q_i,t)-3 = 6 \ \frac{q_i-1}{7-5q_i} \ 
\frac{1}{[1+2 b^2 (5-3q_i) t]^2} = 6 \ 
\frac{q_i-1}{7-5q_i}\ \e{-4b^2 (5-3 q_i)t}_{3/2}~, \quad (q_i<7/5).   
\end{align}

\vskip \baselineskip
\noindent
The equation above indicates that all initial $q$-Gaussians, characterized
by $q_i<7/5$, present a kurtosis that will relax to the standard
Gaussian ($q=1$), following  
$q$-exponentials with the same relaxation index, 
$q_{\rm rel}=3/2$, but different relaxation 
times, $1/[4b^2 (5-3 q_i)]$. Such a dependence of 
the arguments of
these $q$-exponential functions on the initial entropic index $q_i$ 
imply on longer relaxation times for larger values of $q_i$.
This is in agreement with the results of the 
numerical calculation exibited in Fig~\ref{fig:DistrQ1}, corresponding to 
the time evolution of two distributions initialized as $q$-Gaussians,
with $q_i=3/4$ and $q_i=5/4$, respectively.

\begin{figure}[tb]
\centering
\includegraphics[angle=270,width=0.60\textwidth]{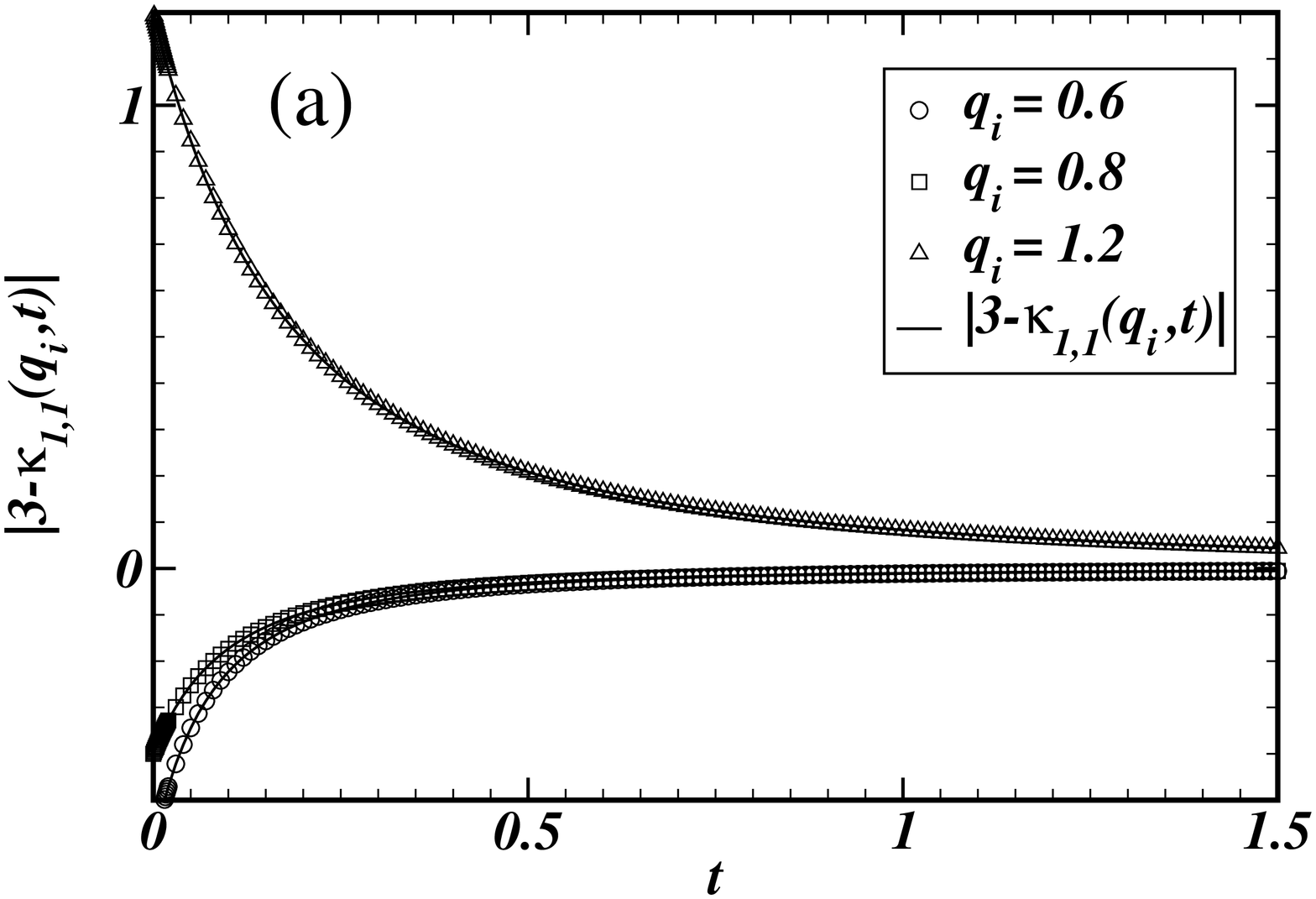}
\includegraphics[angle=270,width=0.60\textwidth]{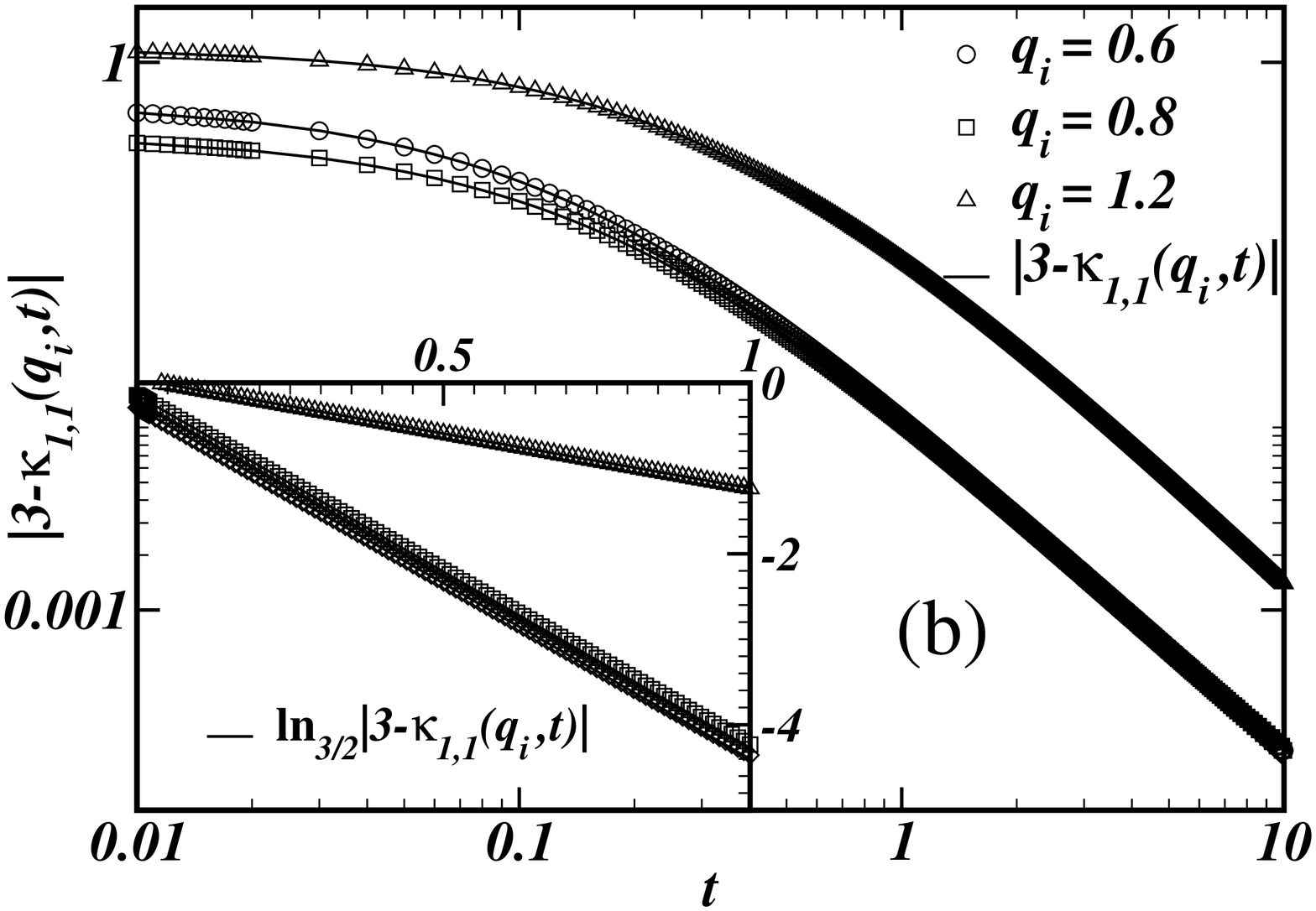}
\caption{Time evolution of the kurtosis obtained by a numerical integration
of the linear diffusion equation 
[\eq{eq:NDiffEq} with $q=1$] is compared to the one of the kurtosis 
$\kappa_{1,1}(q_i,t)$, calculated exactly [\eq{eq:Kurtosis_3}], for typical
values of $q_i$. The data is exhibited in (a) linear-linear, (b) log-log
and $q$-logarithm-linear [inset of figure (b)] plots. The straight lines in
the inset of figure (b) [notice that the data for $q_i=0.6$ and $q_i=0.8$
appear essentially superposed] ensure the index $q_{\rm rel}=3/2$ of 
the relaxation
process, and their slopes yield the corresponding relaxation times.}
\label{fig:Lin_kurt}
\end{figure}
 
As a test for our numerical algorithm, the result of \eq{eq:Kurtosis_3} was
reproduced by a numerical 
integration, for typical values of $q_i$, as shown in Fig~\ref{fig:Lin_kurt}. 
The integration was carried out
using a method based on distributed approximating functionals~\cite{Zhang:97}. 
In all cases, we considered the distribution of \eq{eq:Diff_Ansatz}
with $b_{q_i}(t=0)=1$, as the initial state.  
The results are represented in different scales, like the linear-linear 
[Fig.~\ref{fig:Lin_kurt}(a)] and double logarithm
[Fig.~\ref{fig:Lin_kurt}(b)] ones. However, an elegant way to show that
$\kappa_{1,1}(q_i,t)$ decreases as a $q$-exponential function, with 
$q_{\rm rel}=3/2$,
is by representing the data in terms of the inverse function, i.e., 
the corresponding $q$-logarithmic function. This is exhibited in the
$q$-logarithm-linear plots in the inset of Fig.~\ref{fig:Lin_kurt}(b), 
where one observes perfect linear fits, whose slopes give 
the associated relaxation times, $1/[4(5-3q_i)]$. 

\begin{figure}[tb]
\centering
\includegraphics[angle=270,width=0.60\textwidth]{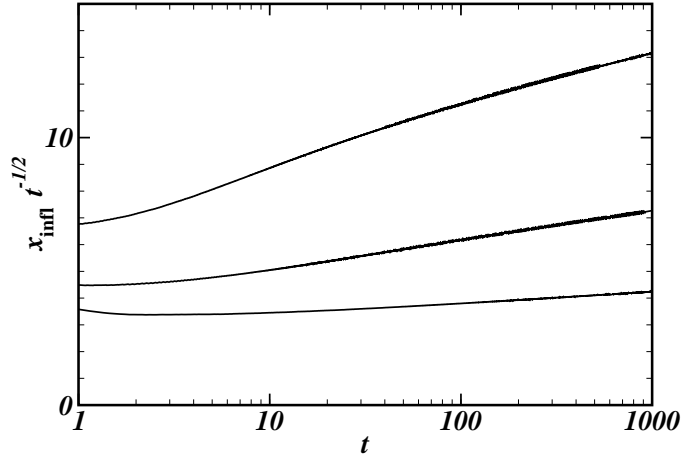}
\caption{Time evolution of the inflection-point
position, $x_{\rm infl}$ (scaled by $\sqrt{t}$), for typical initial
$q$-Gaussians, characterized respectively, by 
$q_{i}=1.2, 1.5, \ {\rm and} \ 2.0$ (from top to bottom).
The small fluctuations are just a numerical artifact.}
\label{fig:infpoint}
\end{figure}

In a recent analytical work it was shown that the
diffusion equation, when initialized with a $q$-Gaussian distribution
($1<q_i<3$), will asymptotically approach its final solution, i.e., 
a Gaussian distribution ($q=1$)~\cite{Anteneodo:06}. Herein, we present a
numerical method to investigate how this change occurs 
for different values of $q_i$ within this range.
One may see easily that the $q$-Gaussian distributions, defined in 
\eq{eq:Diff_Ansatz}, present an inflection point for $q>1$, when
represented in a semi-logarithmic plot; the same does not
occur for $q \leq 1$. Therefore, one
expects that in the transformation process from  
an initial $q$-Gaussian, characterized by an entropic index $q_i>1$, 
to the
asymptotic Gaussian, with $q=1$, such an inflection point should remain
at intermediate times, i.e., in the transient regime, and afterwards, it
should approach infinity.  
An example of this effect is shown in the inset of 
Fig.~\ref{fig:DistrQ1}(b), where we exhibit the corresponding time
evolution of an initial $q$-Gaussian distribution denoted by 
$q_i=1.25$. 
Hence, the time evolution of this inflection point may provide some
additional information regarding the process of approach to the Gaussian
distribution, and in particular, in the cases $q_i>7/5$, for which the
kurtosis of \eq{eq:Kurtosis_3} is not defined. 
Let us herein denote the
position of the inflection point by $x_{\rm infl}$; we analyze the time
evolution of the rescaled quantity, $(x_{\rm infl}/\sqrt{t})$, in order to
measure the time evolution of the inflection point taking off the usual
spreading effect of the distribution during the diffusion process.
We have followed the time evolution of $(x_{\rm infl}/\sqrt{t})$ 
for different initial
values $q_i$, as shown in Fig.~\ref{fig:infpoint}. Within the time interval
feasible for computational purposes, we have noticed that 
$(x_{\rm infl}/\sqrt{t})$ always increases in time and hopefully diverges,
in agreement with the 
results of Ref.~\cite{Anteneodo:06}. However, one notices
that for 
higher values of $q_i$, such an increase occurs very slowly, and in
particular, the case $q_i=2$ suggests that the transformation to the
asymptotic Gaussian distribution should take place at a very long
time. The fact that a single diffusing particle, described in terms of a
linear 
equation [\eq{eq:NDiffEq}], may take a very long time to reach its
asymptotic-diffusing regime, supports the result found in some
Hamiltonian systems, described by a set of $N$ coupled
differential equations, for which, given some initial conditions, 
the final equilibrium may never be reached in the thermodynamic limit
($N \rightarrow \infty$)
\cite{LatoraRuffo:98,Anteneodo:98,Moyano:06,Pluchino:07,Pluchino:08,%
nobretsallis:03,nobretsallis:04}. 

\section{General case: ${q}\neq 1$}
\label{sec:qfneq1}

In this section we analyze the general case $q \neq 1$, which corresponds
to the nonlinear porous-medium equation.
The investigation of the solutions of such equation
was done through a numerical integration of 
\eq{eq:NDiffEq}, using the same method applied in the previous section.
Therefore, we followed the time evolution of the kurtosis, which
starts from $\kappa_{\bar{r},q}(q_i,0)$ [\eq{eq:Kurt_gen_s}] and 
will evolve towards its asymptotic limit,  
$\kappa_{\bar{r},q}(q)$ [\eq{eq:kurt_q}]. In particular, we will search
for the relaxation law associated with  
$|\kappa_{\bar{r},q}(q)-\kappa_{\bar{r},q}(q_i,t)|$. 

\begin{figure}[tb]
\centering
\includegraphics[angle=270,width=0.45\textwidth]{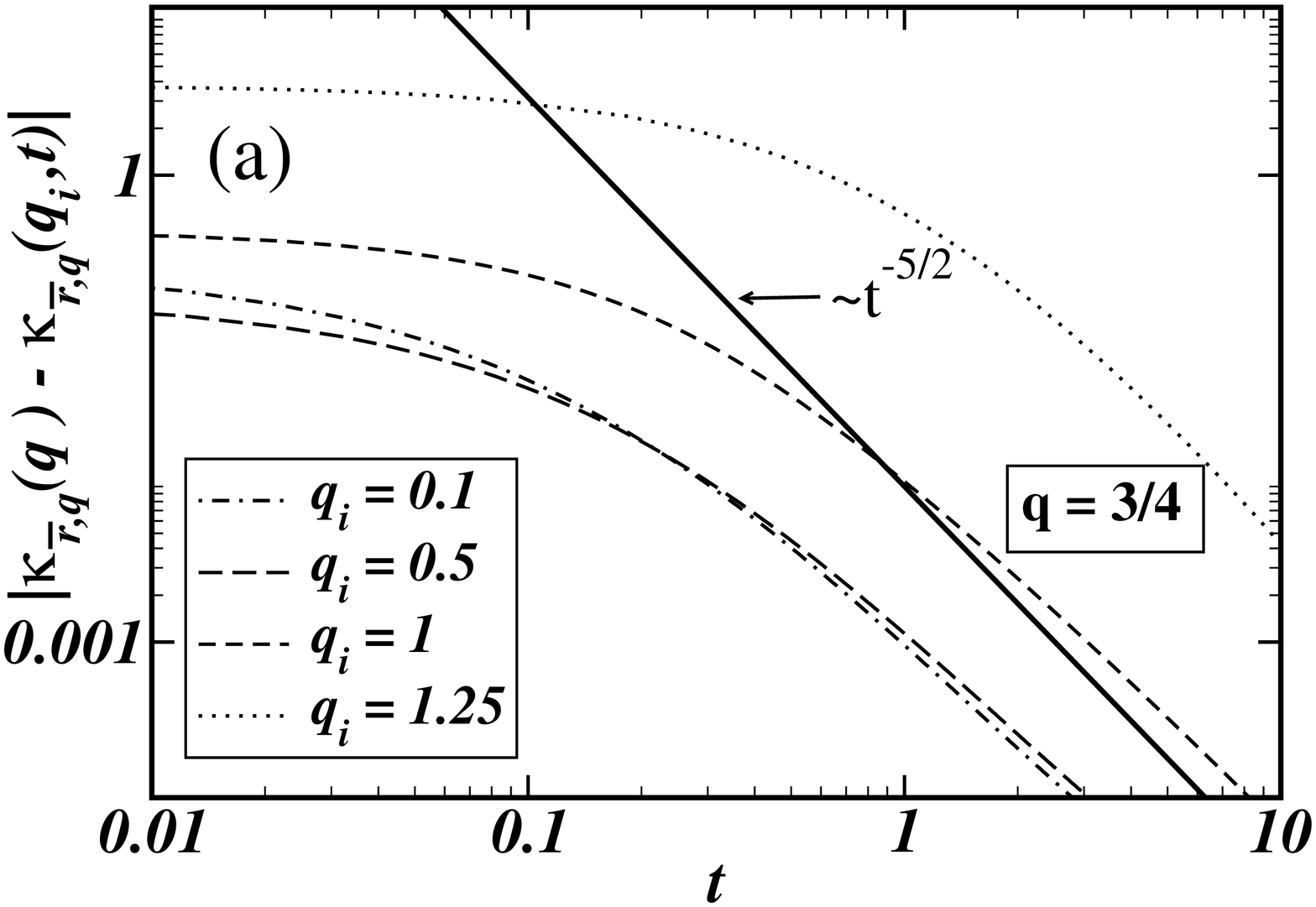}
\hspace{1.0cm}
\includegraphics[angle=270,width=0.45\textwidth]{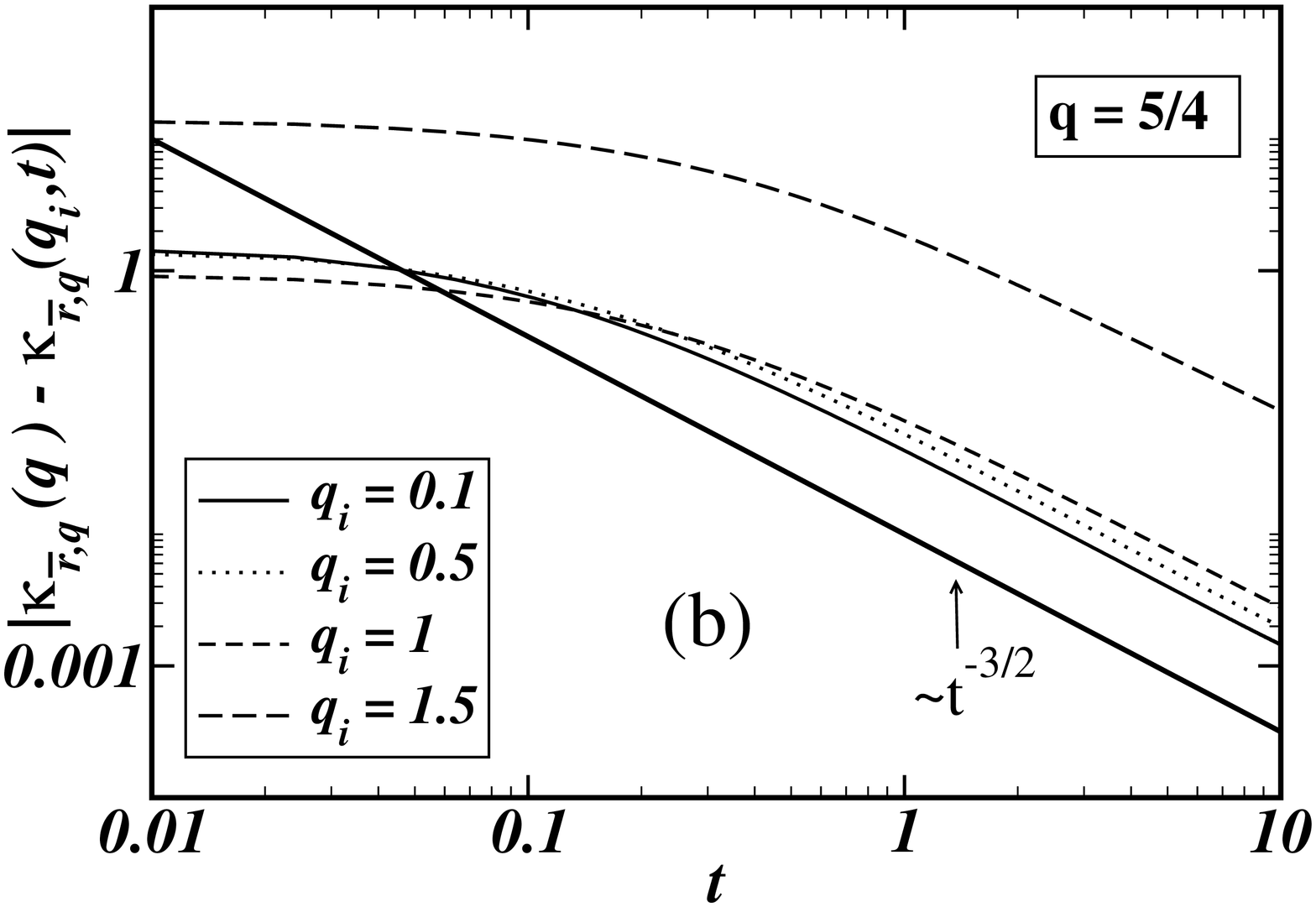}
\includegraphics[angle=270,width=0.45\textwidth]{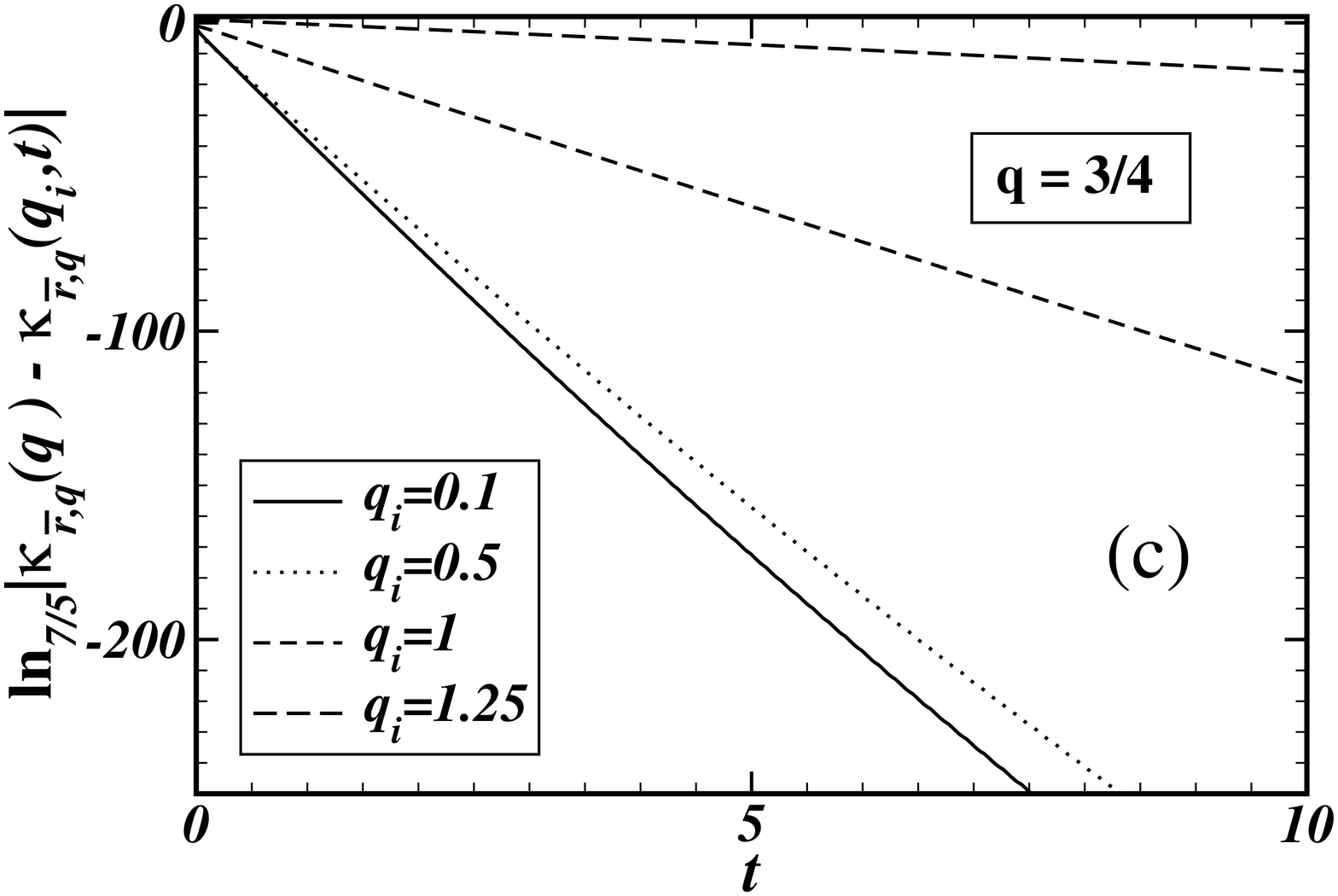}
\hspace{1.0cm}
\includegraphics[angle=270,width=0.45\textwidth]{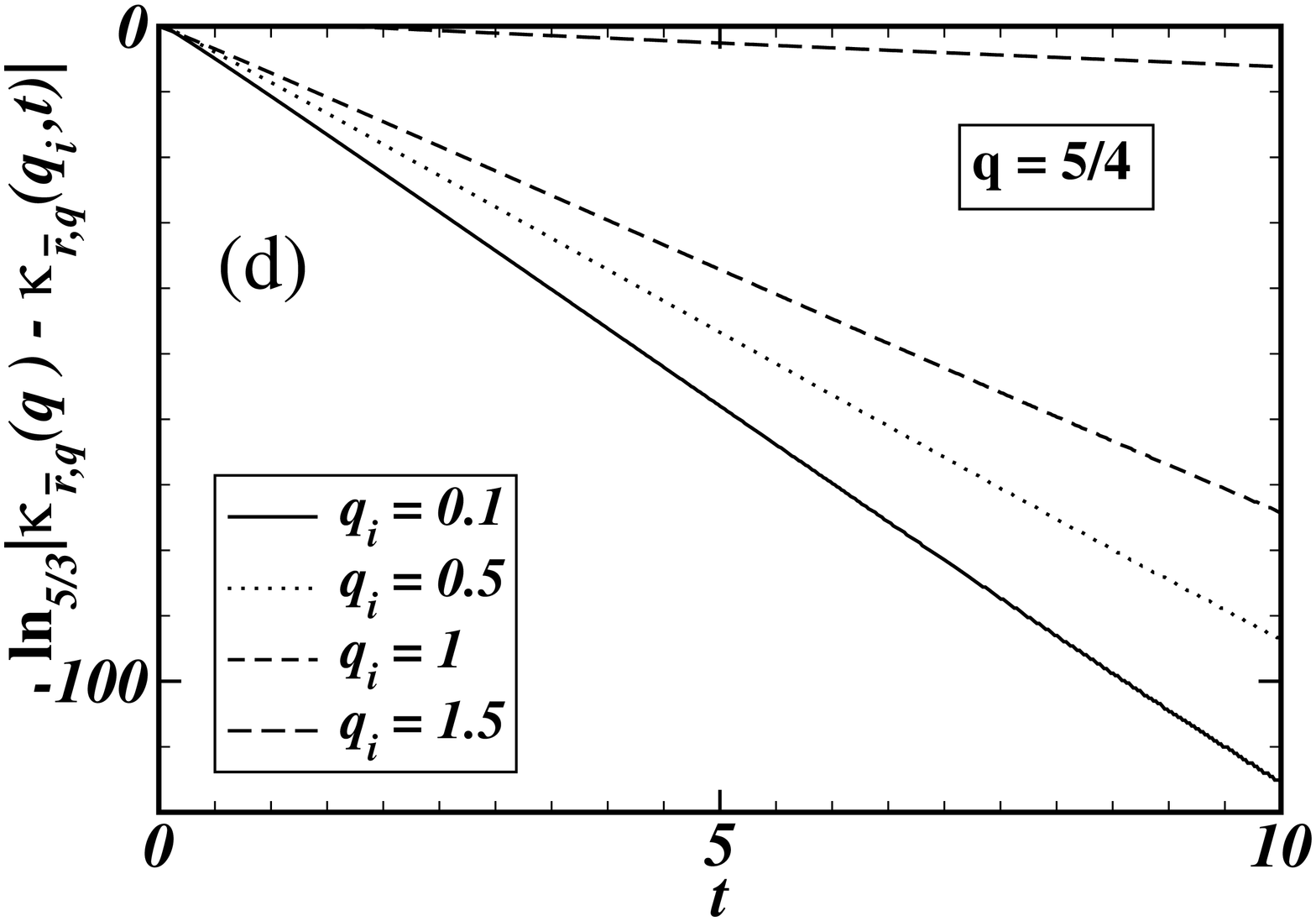}
\caption{Time evolution of the absolute values of the 
differences between the kurtosis at their initial and final 
values, 
$\kappa_{\bar{r},q}(q_i,t)$ [\eq{eq:Kurt_gen_s}] and 
$\kappa_{\bar{r},q}(q)$ [\eq{eq:kurt_q}], respectively, for typical
initial and final $q$-Gaussians. 
(a) $q=3/4$ and several initial values
of $q_i$; the full straight line corresponds to the power-law decay
$t^{-5/2}$. 
(b) $q=5/4$ and several initial values
of $q_i$; the full straight line corresponds to the power-law decay
$t^{-3/2}$. 
(c) The data of (a) is represented in a $\log_{7/5}$ scale.
(d) The data of (b) is represented in a $\log_{5/3}$ scale.}
\label{fig:Kurt0.75_1.5}
\end{figure}

In Fig.~\ref{fig:Kurt0.75_1.5} we exhibit the quantity 
$|\kappa_{\bar{r},q}(q)-\kappa_{\bar{r},q}(q_i,t)|$  
for two typical
final $q$-Gaussians, namely $q=3/4$ and $q=5/4$, starting the
numerical procedure with different initial $q$-Gaussians. Similarly to what
happened in the previous section, our numerical investigation yields, 
in both cases, that the kurtosis relaxes to the corresponding final values,
$\kappa_{\bar{r},q}(q)=2.34797...$ $(q=3/4)$ and 
$\kappa_{\bar{r},q}(q)=3.81717...$ $(q=5/4)$, 
according to $q$-exponentials,
whose relaxation index $q_{\rm rel}$ depends only on $q$. 
The full straight lines in Figs.~\ref{fig:Kurt0.75_1.5}(a) and 
\ref{fig:Kurt0.75_1.5}(b) correspond to the power-law decays, 
$t^{-5/2}$ and $t^{-3/2}$, which are associated with
$q$-exponentials characterized by the relaxation indexes,   
$q_{\rm rel}=7/5$ (for $q=3/4$) and $q_{\rm rel}=5/3$ 
(for $q=5/4$), respectively.
Taking into
account the results obtained in the previous section as well, i.e., a
relaxation following a $q$-exponential with 
$q_{\rm rel}=3/2$ (for $q=1$), we propose a general form for the 
relaxation of the kurtosis,  

\begin{equation}
\label{eq:relaxqiqf}
|\kappa_{\bar{r},q}(q)-\kappa_{\bar{r},q}(q_i,t)| =  
A(q_i,q) \left[ 1 + (1-q_{\rm rel})b^{2}f(q_i,q) t 
\right]^{1/(1-q_{\rm rel})}~, 
\end{equation}

\vskip \baselineskip
\noindent
where the index $q_{\rm rel}$ associated with the $q$-exponential 
relaxation process
depends only on the index $q$, characteristic of the asymptotic
$q$-Gaussian distribution, and that it appears to follow the heuristic
relation, 

\begin{equation}
\label{eq:qversusqf}
q_{\rm rel}(q) = \frac{2q-5}{2q-4}~.
\end{equation}

\begin{figure}[tb]
\centering
\includegraphics[angle=270,width=0.60\textwidth]{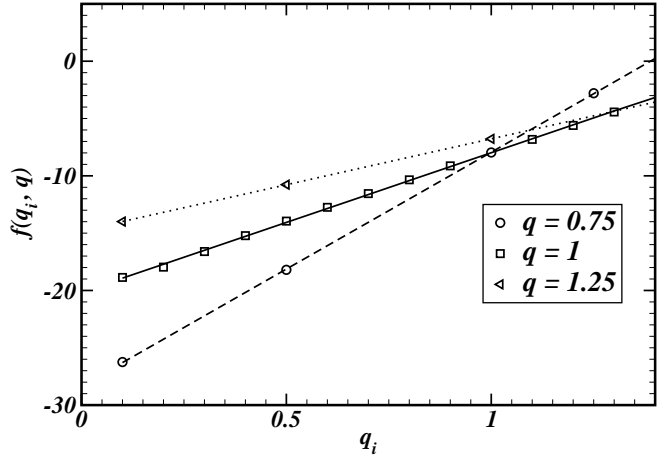}
\caption{The argument $f(q_i,q)$ of the $q$-exponential 
characterizing the relaxation
of the kurtosis, from an initial $q$-Gaussian (index $q_i$) 
to an asymptotic $q$-Gaussian (index $q$), as defined in
\eq{eq:relaxqiqf}, is exhibited as a function of $q_i$, for different
values of $q$. In the case $q=1$, the points (represented by squares)
were computed numerically,
whereas the full line corresponds to the analytical result. In the case
$q=0.75$ ($q=1.25$) the points were computed numerically, and the
straight dashed (dotted) line corresponds to a linear fit.}
\label{fig:Kurt_f}
\end{figure}

\vskip \baselineskip
\noindent
The coefficient $A(q_i,q)$ that appears in \eq{eq:relaxqiqf}
should satisfy $A(q,q)=0$, in such a way that starting 
the numerical integration procedure with the 
exact solution of \eq{eq:NDiffEq}, the kurtosis should not change in time,
i.e., the initial 
distribution remains stable for all times. The argument of the
$q$-exponential, $f(q_i,q)$, should recover the particular case 
$q=1$, calculated analytically,
$f(q_i,1)=-4(5-3q_i)$ [cf. \eq{eq:Kurtosis_3}]. This argument
was estimated numerically
for typical values of $q \neq 1$, as shown in 
Fig.~\ref{fig:Kurt_f}, and our results suggest a simple (essentially
linear) general form, $f(q_i,q)=a(q)+b(q)q_i$. 

\section{Conclusions}

We have analized, using both analytical and numerical approaches, 
the stability of $q$-Gaussian distributions as particular solutions of the
porous-medium equation. This was done by investigating the relaxation
towards the final, asymptotic $q$-Gaussian solution, 
characterized by an index $q_f$, when considering as an initial
distribution a $q$-Gaussian,  
specified by an index $q_i$. By following the time evolution of the
kurtosis (defined for $q_i<7/5$, but always finite in the asymptotic 
limit, $ t \gg 1$), we have found evidence that such a relaxation process
follows a $q$-exponential function, characterized by a relaxation index
$q_{\rm rel}(q)$. 
Therefore, in principle, the problem considered may be formulated in terms
of four indexes:
(i) the index $q$ defined by the porous-medium equation; 
(ii) $q_{i}$, 
associated with the initial $q$-Gaussian distribution; (iii) $q_{f}$,
associated with the final $q$-Gaussian distribution; 
(iv) $q_{\rm rel}$, related to the $q$-exponential of the 
relaxation towards the asymptotic distribution. 
Since we have found no evidence of $q_f \neq q$, we assumed that
$q_f \equiv q$; this supposition is also supported by a recent analytical
approach of the asymptotic behavior of the linear diffusion equation
\cite{Anteneodo:06}. Accordingly, our study was restricted  
to three indexes, namely, $q, \, q_i$, and $q_{\rm rel}$. 

In the definition of this kurtosis, the powers ($s$ and $r$) that 
appear in the probability distributions of the second and fourth 
generalized moments were chosen conveniently in order to
yield a finite kurtosis in the limit $t \gg 1$; although these choices are
arbitrary, we expect that other alternatives (e.g., those used in
Ref.~\cite{Estrada:08}) should not change the present results
qualitatively. 
By using a
numerical approach based on the evolution of the inflection point that
appears in a semi-logarithmic plot of a $q$-Gaussian with $q>1$, we have
observed that in some cases, an initial infinite-variance    
distribution ($q_i \ge 5/3$) may take a very long time to be transformed
into  
a finite-variance one ($q<5/3$). In particular, considering the linear
diffusion equation, we have shown through this method that an
infinite-variance distribution  
($q_i \ge 5/3$) evolves very slowly in time towards the asymptotic Gaussian
distribution. 
The fact that a single diffusing particle, described in terms of a
linear 
equation, may take a very long time to reach its
asymptotic-diffusing regime, supports the existence a metastable state 
found in some highly-interacting
Hamiltonian systems, described by a set of $N$ coupled linear
differential equations, whose duration diverges in the thermodynamic limit  
($N \rightarrow \infty$). 
Moreover, for a finite (but sufficiently large) $N$, it has been found
recently that
the angles described by the infinite-range-interaction
XY rotator model follow a distribution 
\emph{in the long-time-limit} (i.e., in the limit for which its
kinetic temperature coincides with the one of the BG canonical ensemble)
that is 
well-fitted by a $q$-Gaussian, with $q\approx1.5$~\cite{Moyano:06}. This
suggests that in such Hamiltonian models the BG equilibrium state 
is approached through different 
steps, one of them being the attainment of the equilibrium temperature; 
in the simpler system considered herein, the approach to the final,
asymptotic solution, follows a relaxation behavior that may be also 
very slow in some cases. 

\noindent
\acknowledgments
{C. T. is grateful to C. Anteneodo and R. S. Mendes for fruitful 
discussions. All authors thank the
Brazilian agencies CNPq, Faperj and Pronex for financial support.}

\bibliographystyle{revtex}
\bibliography{entropy}

\end{document}